\documentclass[prd,nofootinbib,preprintnumbers,twocolumn,showpacs]{revtex4}
\usepackage{graphicx,epsf,epsfig}
\usepackage{amsmath}
\usepackage{color}
\usepackage[sort&compress]{natbib}
\usepackage{amssymb}
\usepackage{bbm}
\usepackage{tabls,multirow}

\renewcommand{\arraystretch}{1.2}

\def\vev#1{\left\langle #1\right\rangle}

\newcommand{\be}{\begin{equation}}
\newcommand{\ee}{\end{equation}}
\newcommand{\bd}{\begin{displaymath}}
\newcommand{\ed}{\end{displaymath}}
\newcommand{\bea}{\begin{eqnarray}}
\newcommand{\eea}{\end{eqnarray}}
\newcommand{\nn}{\nonumber}

\begin{document}
\title{Flavour structure of supersymmetric $SO(10)$ GUTs with extended matter sector}
\author{Martin Heinze}\email{mheinze@kth.se}
\affiliation{
Department of Theoretical Physics, School of Engineering Sciences,  
Royal Institute of Technology (KTH), AlbaNova University Center, Roslagstullsbacken 21, SE-106 91 Stockholm, Sweden 
}
\author{Michal Malinsk\'{y}}\email{malinsky@ific.uv.es}
\affiliation{
Department of Theoretical Physics, School of Engineering Sciences, 
Royal Institute of Technology (KTH), AlbaNova University Center, Roslagstullsbacken 21, SE-106 91 Stockholm, Sweden 
}
\affiliation{AHEP Group, Instituto de F\'{\i}sica Corpuscular -- C.S.I.C./Universitat de Val\`encia, Edificio de Institutos de Paterna, Apartado 22085, E 46071 Val\`encia, Spain}
\begin{abstract}
We discuss in detail the flavour structure of the supersymmetric $SO(10)$ grand unified models with the three traditional 16-dimensional matter spinors mixed with a set of extra 10-dimensional vector multiplets which can provide the desired sensitivity of the SM matter spectrum to the GUT symmetry breakdown at the renormalizable level.   
We put the qualitative argument that a successful fit of the quark and lepton data requires an active participation of more than a single vector matter multiplet on a firm, quantitative ground. We find that the strict no-go obtained for the fits of the charged sector observables in case of a single active matter 10 is relaxed if a second vector multiplet is added to the matter sector {and excellent, though non-trivial, fits can be devised}.  Exploiting the unique calculable part of the neutrino mass matrix governed by the $SU(2)_{L}$ triplet in the $54$-dimensional Higgs multiplet, {a pair of genuine predictions of the current setting is identified: a non-zero value of the leptonic 1-3 mixing close to the current $90\%$ C.L. limit and a small leptonic Dirac CP phase are strongly preferred by all solutions with the global-fit $\chi^{2}$-values below 50.} \end{abstract}
\pacs{12.10.-g, 12.15.Ff, 12.60.Jv, 14.60.Pq}
\maketitle
\section{Introduction\label{sect:introduction}}
Even after 35 years since the pioneering work by Georgi and Glashow \cite{Georgi:1974sy} the idea of grand unification still receives a lot of attention across the high energy physics community, providing one of the most popular schemes beyond the Standard Model (SM) of particle interactions. Apart from the canonical prediction of the proton instability and monopoles, the simplest grand unified theories (GUTs) can be tested for the compatibility between the observed SM flavour structure and the simplified shape of their Yukawa sector emerging at the grand unification scale $M_{G}$, typically in the ballpark of $10^{16}$~GeV. 

Recently, with the advent of the precision neutrino physics \cite{Strumia:2006db}, 
the field experienced a further renaissance fuelled by the observation of neutrino flavour oscillations \cite{Schwetz:2008er}. The eV scale of the light neutrino masses governing  these phenomena is often connected to the scale of the new physics underpinning a variant of the seesaw mechanism \cite{seesaw}. 
To this end, GUTs can provide a very detailed information on the relevant high energy dynamics, with implications for the position of the seesaw thresholds and, hence, the absolute neutrino mass scale.

With the new piece of information at hand, the flavour structure of the simplest GUTs has been scrutinized thoroughly in the past \cite{flavour-in-SO10-elder,
flavour-in-SO10-newer,
flavour-in-SU5}. 
The intriguing pattern of flavour mixing in the lepton sector, together with the constraints on the absolute neutrino mass scale, turned out to be extremely useful in discriminating among the simplest potentially realistic GUTs, in particular those based on the $SO(10)$ gauge symmetry \cite{Carlson:1975gu}. 

The main virtue of the $SO(10)$ framework consists in the fact that every SM matter generation fits perfectly into a single 16-dimensional chiral spinor of $SO(10)$, thus providing a simple rationale for the very special anomaly-free pattern of the SM hypercharges. On top of that, the right-handed neutrino is inevitable and, hence, seesaw is naturally accommodated. As a rank=5 gauge symmetry, $SO(10)$ also admits a large number of viable symmetry breaking chains 
\cite{SO10chains}, 
resulting in many different intermediate scale scenarios with rich phenomenology. 

From the neutrino perspective, the most important aspect of this freedom is the scale of the $B-L$ symmetry breakdown. In the most popular schemes it is triggered either by the vacuum expectation values (VEVs) in the $16$-dimensional $SO(10)$ spinors or in the irreducible components of the five-index antisymmetric tensor ($126\oplus \overline{126}$) in the Higgs sector. 
In supersymmetric (SUSY) scenarios with $126_{H}\oplus \overline{126}_{H}$, the $R$-parity of the minimal supersymmetric standard model (MSSM) emerges naturally as a remnant of the $SO(10)$ gauge symmetry~\cite{naturalRparity,Aulakh:2000sn}, 
there are no proton-dangerous $d=4$ operators and a potentially realistic Yukawa sector with a calculable seesaw can be implemented at the renormalizable level \cite{minimalSO10,renormalizableseesaw}.

On the other hand, one has to resort to a cumbersome Higgs sector as further multiplets are needed to break through an intermediate $SU(5)$ symmetry which is left intact by the SM singlets in $126_{H}\oplus \overline{126}_{H}$. 
Remarkably, none of the simplest options, i.e., neither extra 45 nor 54, is sufficient to do so at the renormalizable level\footnote{In this respect, the situation in the non-supersymmetric setting differs substantially from the supersymmetric case, c.f. \cite{Bertolini:2009es}.} \cite{Aulakh:2000sn}, and even with both of them, non-renormalizable operators are still needed to mix the $SU(2)_{L}$-doublets in $\overline{126}_{H}$ with those from other Higgs multiplets ($10$- or $120$-dimensional) in order to get a reasonable Yukawa sector. Actually, renormalizability calls for $210_{H}$ instead which can provide both the $SU(5)$ breakdown as well as the doublet mixing, yet retaining a high level of predictivity.
Unfortunately, the minimal renormalizable SUSY $SO(10)$ model \cite{minimalSO10} 
with $10_{H}\oplus 126_{H}\oplus \overline{126}_{H}\oplus 210_{H}$ in the Higgs sector does not seem to work due to the generic tension between the neutrino mass scale and SUSY unification constraints \cite{killingminimalSO10}. 
Recently, there have been several attempts to overcome this issue by, e.g., invoking split SUSY \cite{Bajc:2008dc} or by employing a $120$-dimensional Higgs representation (see for instance \cite{Bertolini:2004eq,SO10with120} 
and references therein).   
However, most of these constructions are plagued by the instability of the perturbative description due to a Landau pole emerging close to the GUT scale \cite{Barr:1981qv,Aulakh:2000sn}.

The situation in models with $16_{H}\oplus \overline{16}_{H}$ triggering the $B-L$ breakdown \cite{so10with16} 
is quite different in several aspects. First, a concise Higgs sector of the form $16_{H}\oplus \overline{16}_{H}\oplus 45_{H}\oplus 54_{H}$ is sufficient to break through the $SU(5)$ lock\footnote{This statement, however, is not a trivial analogue of a similar mechanism at play in the $126\oplus \overline{126}$ case because the product $16_{H} 54_{H} \overline{16}_{H}$, unlike  $126_{H} 54_{H} \overline{126}_{H}$, does not contain a gauge singlet and thus one of the parameters is missing here.}. Second, here there is no problem with mixing the $SU(2)_{L}$-doublets in $10_{H}$ (which is again introduced for the sake of a potentially realistic Yukawa sector) with those in $16_{H}\oplus \overline{16}_{H}$ at the renormalizable level.  Moreover, the Landau pole is safely postponed beyond the Planck scale. 

In spite of these attractive features, it turns out to be rather difficult to construct predictive models along these lines in practise. The basic reason is that there is no way to communicate the information about the $SU(2)_{R}\otimes U(1)_{B-L}$ and $SU(5)$ symmetry breaking (driven by the VEVs of $16_{H}\oplus \overline{16}_{H}$ and $45_{H}\oplus 54_{H}$) to the matter sector spinorial bilinears $16_{M}16_{M}$ at the renormalizable level.  Thus, in order to get potentially realistic effective quark and lepton spectra and mixings, non-renormalizable operators must be invoked and there is a need for further assumptions to retain predictivity in the Yukawa sector, see, e.g., \cite{flavour-in-SO10-elder} 
and references therein.  

An elegant solution \cite{extra10matter} 
to this conundrum consists in abandoning the ``matter in spinors'' paradigm of the $SO(10)$ model building. With extra 10-dimensional $SO(10)$ matter vectors in the game (to be denoted by $10_{M}$) admixing at a certain level into the light matter fields, the basic invariants of the form $16_{M}10_{M}16_{H}$, $10_{M}10_{M}54_{H}$ and $10_{M}10_{M}45_{H}$ do the magic at the renormalizable level. Moreover, since the SM-singlet VEV of $16_{H}$, $\vev{16_{H}}$, governing the mixing between the spinors and vectors can be comparable to the scale of the (gauge singlet) mass term $M_{10}10_{M}10_{M}$, the matter vectors do not need to decouple from the electroweak-scale ($v$) physics - it's not the $v$ over $M_{10}$ but the $\vev{16_{H}}$ over $M_{10}$ ratio that matters. 

This is even more so in the SUSY GUTs where a single-step breaking (bringing $\vev{16_{H}}$ to the vicinity of the GUT scale $M_{G}$) is typically favoured. Furthermore, if one admits a hierarchy in the eigenvalues of even a Planck-scale $M_{10}$ that could originate from a similar source like, e.g., the hierarchy of the Yukawa couplings, it's very plausible to expect at least one of them at around (or even below) $M_{G}$. This, indeed, makes observable non-decoupling effects of the extra $10_{M}$'s very natural. Remarkably, in such a case, the relative magnitude of the $SU(2)_{R}\otimes U(1)_{B-L}$ and $SU(5)$ breaking observed in the MSSM matter spectra (of the order of the differences in the second to third generation mass ratios, i.e., few percent) is nicely linked to the hierarchy of the SUSY GUT-scale thresholds. Moreover, the triplet contribution to the neutrino mass matrix turns out to be calculable in this framework because the leptonic $SU(2)_{L}$ doublets in $10_{M}$ can couple to the Higgs triplet in $54_{H}$ at the renormalizable level. 

Let us also note that the extra vectors in the matter sector are inevitable in the unified models beyond $SO(10)$, like, e.g., in $E_{6}$ GUTs \cite{E6}. 
Recently, the extra matter in the $SO(10)$ GUT context played a central role in works \cite{gaugemediation} 
in which a class of phenomenologically viable  models of tree-level gauge mediation as means of SUSY breaking has been constructed.    

Although this framework has been used before by several authors 
to {address, e.g., the flavour problem of the SM or to constrain the SUSY flavour and CP structure of its GUT-inspired extensions \cite{vectormatterused},  a generic} study of the flavour structure of the SUSY $SO(10)$ GUTs with vector multiplets in the matter sector has been carried out only partially, namely for a single vector matter multiplet at play in \cite{Malinsky:2008zz} where a no-go for the simplest setting has been formulated. 
In this study we attempt to go beyond the minimal case and look at the viability of a more realistic scenario in which a hierarchy in the Planck-scale $M_{10}$ brings a pair of its eigenvalues to the vicinity of the GUT scale. As we shall see, the generic no-go of \cite{Malinsky:2008zz} is lifted already for the second lightest eigenvalue of $M_{10}$ contributing with just around $1\%$ of the strength of the first one and, {even within such a 
``quasi-decoupled'' setting,} the flavour structure of the SM charged matter sector is accommodated in a very natural manner.

{Remarkably, complete fits including the triplet-dominated neutrino sector observables require a significant \hyphenation{sig-ni-fi-cant}contribution from the second $10_{M}$ in the matter sector, far from the quasi-decoupled regime. In such a case, the minimality of the Higgs potential is fully exploited and two generic predictions of the scheme can be identified: the best fits of all the measured quark and lepton flavour parameters strongly favour small but non-zero value of the leptonic reactor mixing angle $\theta_{13}^{l}$ within the ballpark of the {current global upper limit \cite{theta13}}, together with a close-to-zero value of the leptonic Dirac CP phase. 
}

The work is organized as follows: In section II we define the basic framework, derive the effective mass matrices for the MSSM matter fields and comment on the role the calculable triplet contribution plays in the neutrino mass matrix. After a brief recapitulation of the no-go for the minimal setting, these formulas are subject to a thorough numerical analysis in section III for the case of a pair of non-decoupled $10_{M}$'s and we comment on the blindness of the best $\chi^{2}$ fits to the contributions associated to the Yukawa coupling of $45_{H}$ {observed in a large part of the parametric space available to good charged sector fits.}  In section IV, we briefly comment on the prospects of a realistic model building and its basic strategies. Then we conclude.
\section{The framework}
Let us begin with a definition of the {minimal framework} in which the generic principles advocated above can be implemented in a potentially viable manner. Since the details of the matter sector flavour structure depend only loosely on the specific shape of the Higgs sector, we shall focus on the simplest conceivable model. The following discussions can be then extended to more complicated settings in a straightforward way. In order to keep the discussion compact, we shall stick to salient points only and, whenever appropriate, refer to work \cite{Malinsky:2008zz} where a similar construction has been discussed in great detail.
 
\subsection{The model definition}  

\subsubsection{The matter sector}
We shall consider the standard three copies of the $SO(10)$ spinors $16_{M}^{i}$ ($i=1,2,3$) {in the matter sector} (otherwise one could not accommodate properly the three generations of up-type quarks), together with $n$ copies of the $SO(10)$ vectors $10_{M}^{k}$ $(k=1,..,n)$.  
{The subscript $M$ indicates that these multiplets are odd under a $Z_{2}^{M}$ matter parity invoked in order to prevent the classical trouble with the $d=4$ proton decay due to their potential mixing with the $Z_{2}^{M}$-even Higgs multiplets carrying a generic subscript $H$.}
The effective matter sector spanned non-trivially over both {$16_{M}$'s and $10_{M}$'s} then exhibits a full sensitivity to the GUT-scale VEVs, overcoming the ``high-energy blindness'' of the purely spinorial matter in the renormalizable settings with $16_{H}\oplus\overline{16}_{H}$.  

The $SU(3)_{c}\otimes SU(2)_{L}\otimes U(1)_{Y}$ structure of these
multiplets reads (in the $Q=T_{L}^{3}+Y$ convention):
\bea\label{decs} 
16_M & =& (3,2,+\tfrac{1}{6})\oplus{(1,2,-\tfrac{1}{2})}\oplus(\overline{3},1,-\tfrac{2}{3}) \\
& \oplus & (\overline{3},1,+\tfrac{1}{3})\oplus(1,1,+1)\oplus{(1,1,0)}\nn
\\
10_M &= & (3,1,-\tfrac{1}{3})  \oplus  {(1,2,+\tfrac{1}{2})}\oplus (\overline{3},1,+\tfrac{1}{3})\oplus{(1,2,-\tfrac{1}{2})}\nn
\eea
The SM sub-multiplets of $16_{M}$ above will be, from now on, consecutively called $Q_L$, $L_L$,
$U^c_L$, $D^c_L$, $N^c_L$  and  $E^c_L$, while those of $10_{M}$ as
 $\Delta_L$, $\Lambda^c_L$, $\Delta^c_L$ and  $\Lambda_L$.

Let us reiterate that at the $SU(3)_{c}\otimes SU(2)_{L}\otimes U(1)_{Y}$ level $D^{c}_{L}$ can mix with $\Delta^{c}_{L}$ and $L_{L}$ with $\Lambda_{L}$ giving rise to the physical down quark and charged-lepton components  (to be called $d^{c}_{L}$ and $l_{L}$), sharing the features of both $16_{M}$ and $10_{M}$, in particular their sensitivity to the {GUT-scale physics}). 

Let us also note that the matter sector spanned on $16_{M}$'s and  $10_{M}$'s  can be viewed as a hint of an underlying $E_{6}$ gauge structure where these multiplets both fit into its fundamental $27$-dimensional representation (decomposing under $SO(10)$ as $27=16\oplus 10\oplus 1$). On the other hand, this correspondence is  rather loose here as we do not demand the number of $10_{M}$'s to match the number of $16_{M}$'s, let alone the absence of the  extra singlets, c.f. section~\ref{sect:massmatrices}.  

\subsubsection{The Higgs sector}
Concerning the Higgs model that can support the desired $SO(10)\to SU(3)_{c}\otimes SU(2)_{L}\otimes U(1)_{Y}$ symmetry breaking chain at the renormalizable level,  the simplest such setting in the SUSY context corresponds to the $16_{H}\oplus\overline{16}_{H}\oplus 45_{H}\oplus 54_{H}$ Higgs sector. Note that $45_{H}$ alone is not enough because the $F$-flatness aligns its VEVs with the SM singlets in $16_{H}\oplus \overline{16}_{H}$ leaving $SU(5)$ unbroken \cite{Aulakh:2000sn}. 

The relevant factors consist of the following SM components:
\bea
16_H & =& (3,2,+\tfrac{1}{6})\oplus{(1,2,-\tfrac{1}{2})}\oplus(\overline{3},1,-\tfrac{2}{3})\nn \\
& \oplus & (\overline{3},1,+\tfrac{1}{3})\oplus(1,1,+1)\oplus\underline{(1,1,0)}\nn\,,
\\
\overline{16}_H & =& (\overline{3},2,-\tfrac{1}{6})\oplus{(1,2,+\tfrac{1}{2})}\oplus(3,1,+\tfrac{2}{3})\nn \\
& \oplus & (3,1,-\tfrac{1}{3})\oplus(1,1,-1)\oplus\underline{(1,1,0)}\,,
\\
45_H & = & 
(1,3,0)
\oplus (1,1,+1)\oplus\underline{(1,1,0)}\oplus (1,1,-1)\nn\\
& \oplus & (8,1,0)\oplus \underline{(1,1,0)}\oplus(3,1,+\tfrac{2}{3})\oplus (\bar{3},1,-\tfrac{2}{3})\nn\\
& \oplus & (3,2,-\tfrac{5}{6})\oplus(3,2,+\tfrac{1}{6})\oplus (\overline{3},2,+\tfrac{5}{6})\oplus (\overline{3},2,-\tfrac{1}{6})\,,\nn \\
54_H & = & \underline{(1,1,0)}\oplus(1,3,0)\oplus(1,3,+1)\oplus(1,3,-1)\nn \\
&\oplus& (\overline{6},1,+\tfrac{2}{3})\oplus(6,1,-\tfrac{2}{3})\oplus(8,1,0)\oplus (3,2,+\tfrac{1}{6})\nn \\
&\oplus& (3,2,-\tfrac{5}{6}) \oplus (\overline{3},2,-\tfrac{1}{6})\oplus(\overline{3},2,+\tfrac{5}{6})\nn\,,
\eea
where the underlined SM singlets are all expected to receive GUT-scale VEVs. These we shall call $V^{16}$, $V^{\overline{16}}$, $V^{45}_{\Delta}$ (the one in $(15,1,1)_{45}$ with respect to the $SU(4)_{C}\otimes SU(2)_{L}\otimes SU(2)_{R}\subset SO(10)$), $V^{45}_{\Lambda}$ (the one in $(1,1,3)_{45}$ in the same notation) and $V^{54}$, respectively\footnote{Recall that $V^{16}$ is connected to $V^{\overline{16}}$ by the desired $D$-flatness of the SUSY vacuum: $|V^{16}|=|V^{\overline{16}}|$. The notation for the singlets in $45_{H}$ is justified by the observation that $V^{45}_{\Delta}$ can give masses only to the quark-like states in $10_{M}$'s while $V^{45}_{\Lambda}$ enters only the leptonic bilinears. This is  clear from the Pati-Salam decomposition of the $SO(10)$ vector which reads $10=(6,1,1)\oplus (1,2,2)$ where the former factor accommodates $\Delta_L\oplus \Delta^c_L$" while the latter corresponds to $\Lambda_L\oplus \Lambda_L^c$.}.
The ultimate $SU(2)_{L}\otimes U(1)_{Y}\to U(1)_{Q}$ breakdown is then driven by the $SU(2)_{L}$-doublets in $16_{H}\oplus \overline{16}_{H}$ together with a pair of extra copies coming from an additional $SO(10)$-vector Higgs multiplet
\be
10_H  =  (3,1,-\tfrac{1}{3})  \oplus {(1,2,+\tfrac{1}{2})}\oplus (\overline{3},1,+\tfrac{1}{3})\oplus{(1,2,-\tfrac{1}{2})}\nn\,,
\ee
which is added as usual in order to end up with a potentially realistic Yukawa sector. In a self-explanatory notation, we shall use the symbols $v_d^{16}$, $v_u^{\overline{16}}$, $v^{10}_{u}$ and $v^{10}_{d}$  for the corresponding doublet VEVs.  Apart from these, the interplay between the $B-L$ and the electroweak breakdown gives rise to a pair of induced VEVs  on the electrically neutral components of $(1,3,\pm 1)$ of $54_{H}$ (to be called $w_{\pm}$). Subsequently, the renormalizable coupling $10_{M}10_{M}54_{H}$ gives rise to a set of Majorana  entries in the relevant neutrino mass matrix, c.f. section~\ref{sect:matrices}. 
\subsubsection{The renormalizable Yukawa superpotential}
The Yukawa superpotential of the model under consideration reads (with all indices and the Lorentz structure suppressed):
\bea
{W}_Y 
& = & {16}_MY{10}_H{16}_M +{16}_MF{16}_H{10}_M\label{WY}\\
&+&{10}_M(\lambda\,{54}_H\ +\eta\,{45}_H+M_{10}){10}_M \nn\,,
\eea
where $Y$ is a $3\times 3$ complex symmetric Yukawa matrix, $F$ is its $3\times n$ general complex analogue in the mixed $16_{M}$--$10_{M}$ sector and $M_{10}$ and $\lambda$ (and $\eta$) are $n\times n$ complex symmetric (antisymmetric) matrices. 
At the $SU(3)_{c}\otimes U(1)_{Q}$ level, the part of our interest can be written as:
\bea
W_Y &\ni& 
U_L Y U^{c}_Lv^{10}_u + N^{c}_LY N_Lv^{10}_u\label{WYbrokenphase}\\
&+& D_LY D^{c}_Lv^{10}_d + E^{c}_LY E_Lv^{10}_d\nn\\
&+& D_L F \Delta^{c}_Lv_d^{16} + E^{c}_LF \Lambda_L^{-}v_d^{16} + N^{c}_LF \Lambda^{c0}_Lv_d^{16}\nn\\
&+&D^{c}_LF\Delta_LV^{16}  + E_LF\Lambda^{c+}_L V^{16}+ N_LF\Lambda^{c0}_L V^{16} \nn\\
&+&\Lambda_L^{0}\lambda\Lambda_L^{0} w_+ + \Lambda^{c0}_L\lambda\Lambda^{c0}_Lw_- \nn\\
&-& \Delta_L \lambda\Delta^{c}_LV^{54} + \tfrac{3}{2} \Lambda^{c+}_L\lambda \Lambda_L^{-} V^{54}+ \tfrac{3}{2} \Lambda^{c0}_L\lambda \Lambda_L^{0} V^{54}\nn\\
&+&\Delta_L \eta \Delta^{c}_LV^{45}_\Delta  + \Lambda^{c+}_L\eta \Lambda_L^{-} V^{45}_\Lambda+ \Lambda^{c0}_L\eta \Lambda_L^{0} V^{45}_\Lambda  \nn\\
&+& \Delta_L M_{10}\Delta^{c}_L + \Lambda^{c+}_L M_{10}\Lambda_L^{-}+ \Lambda^{c0}_L M_{10}\Lambda_L^{0}\nn,
\eea
where the defining $SU(2)_{L}$ doublets have been broken into their components, i.e., $Q_{L}=(U_{L}, D_{L})$, $L_{L}=(N_{L}, E_{L})$, $\Lambda_{L}=(\Lambda_{L}^{0}, \Lambda_{L}^{-})$ and $\Lambda_{L}^{c}=(\Lambda_{L}^{c+}, \Lambda_{L}^{c0})$.  Wherever possible, we have also absorbed the relevant Clebsch-Gordan coefficients into ${\cal O}(1)$ redefinitions of the independent VEVs and/or couplings, with an important exception at line 6 where the ratio of the Clebsches can not be hidden. This, indeed, is the backdoor through which the desired $SU(5)$ symmetry breaking due to a non-zero $V^{54}$ is transferred into the matter sector. 
\subsection{GUT-scale mass matrices\label{sect:matrices}}
The relevant GUT-scale mass matrices  for the matter fields can be readily read out of eq.~(\ref{WYbrokenphase}):
\bea
M_{u}& = & Y v_{u}^{10}\label{Mu}\,,\\
M_d & = & 
\begin{pmatrix}
Y v^{10}_d & F v_d^{16} \\
F^T V^{16} & M_\Delta
\end{pmatrix}  \label{Md}\;,\\
M_e & = & \begin{pmatrix}
Y v^{10}_d & F V^{16} \\
F^T v_d^{16} & M_\Lambda
\end{pmatrix}  \label{Me}\;,\\
M_\nu&=& \begin{pmatrix}
0 & Y v^{10}_u & 0 & F V^{16} \\
\cdot & 0 & 0 & F v_d^{16}\\
\cdot & \cdot & \lambda w_+ & M _\Lambda\\
\cdot & \cdot & \cdot & \lambda w_-
\end{pmatrix}.\label{Mnu}
\eea
For the first three (Dirac) mass matrices above, the following bases have been used: $(U_{L})(U^{c}_{L})$ for $M_{u}$,  $(D_{L},\Delta_{L})(D^{c}_{L},\Delta^{c}_{L})$ for $M_{d}$ and $(E_{L},\Lambda_{L}^{-})(E^{c}_{L},\Lambda_{L}^{c+})$ for $M_{e}$, respectively. The Majorana mass matrix $M_{\nu}$ has been given in the symmetric basis $(N_{L},N_{L}^{c},\Lambda^{0}_{L},\Lambda^{c0}_{L})$. We have also made use of the symmetry properties of $Y$, $M_{10}$, $\lambda$ and $\eta$ and defined 
\bea
M_{\Delta}&\equiv &M_{10} -\lambda V^{54} + \eta V^{45}_\Delta\,, \label{MDeltaLambda}\\
M_\Lambda &\equiv& M_{10} + \tfrac32\lambda V^{54} - \eta V^{45}_\Lambda\nn.
\eea
Inspecting the matrices above one can appreciate the role of the extra vector multiplets  in propagating the information about the intermediate symmetry breaking into the matter sector:
First, since there are no heavy partners to the up-type quarks available the physical spectrum is determined solely by the spinorial bilinear Yukawa $Y$. 
Second, the hierarchy of the down-type quark spectrum is clearly different from the up-type quarks whenever there is a non-negligible admixture of the $\Delta_{L}^{c}$ components in the light eigenstates. For this to be the case, $|FV^{16}|$ should not be negligible with respect to $M_{\Delta}$.
Third, in order to account for the differences in the down-quark and charged-lepton mass hierarchies it is inevitable to have $M_{\Delta}$ different from {$M_{\Lambda}^{T}$} which can happen only if at least one of the $SU(5)$-breaking VEVs $V^{45}_{\Delta}$, $V^{45}_{\Lambda}$ and/or $V^{54}$ is turned on and it is not screened by the $SO(10)$-singlet mass term $M_{10}$ in (\ref{MDeltaLambda}). Thus, at least some eigenvalues of $M_{10}$ are required to be in the vicinity of the GUT scale. Note that, in spite of the $SO(10)$-singlet nature of $M_{10}$, this can easily be the case if $M_{10}$ happens to exhibit a several-orders-of-magnitude hierarchy as some other Yukawa couplings in the game, in particular $Y\propto M_{u}$. 
 
Note also that there are several interesting formal limits in which the matter spectrum reveals an enhanced symmetry pattern: 
\begin{itemize}
\item Putting $V^{45}_{\Lambda}=V_{\Delta}^{45}$ and $V^{54}$ to zero one has $M_{\Delta}=M_{\Lambda}^{T}$ and thus $M_{d}=M_{e}^{T}$ due to the residual $SU(5)$ gauge symmetry left unbroken by $V^{16}$. 
\item For $M_{10}$ strongly dominating the heavy sector masses, the extra vectors $10_{M}^{k}$ decouple and the sensitivity of the light sector to the intermediate symmetry breaking scales is lost. In this case, all Dirac masses are proportional to each other due to the residual $SU(4)_{C}$  Pati-Salam symmetry exhibited by the matter sector, as expected in all settings with $16_{M}Y16_{M}10_{H}$ alone in the Yukawa sector.
\item For $V^{16}\ll M_{\Delta,\Lambda}$ with $M_{10}$, $V^{54}$ and $V^{45}$ at around the GUT scale the effect of the $SU(4)_{C}$ symmetry breaking becomes observable only in the heavy sector because of the effective suppression of the $FV^{16}$ term linking the GUT-scale VEVs to the light eigenstates. In other words, the vector matter does again decouple from the $SO(10)$ spinors.  
\end{itemize}
These remarks demonstrate clearly the internal consistency of formulas (\ref{Mu})-(\ref{Mnu}).
\subsection{Effective mass matrices}
Below the GUT scale the heavy part of the matter spectrum decouples and one is left with the three standard MSSM families. Their masses and mixings are then dictated by their projections onto the defining basis components $16_{M}^{i}$ and $10_{M}^{k}$, providing the desired sensitivity to the GUT symmetry breakdown in the matter sector. 

In what follows, we shall use the calligraphic symbols  ${\cal 
M}_{f}$ (with $f={u,d,e,\nu}$) for the effective MSSM mass matrices to make a clear distinction between these and the full-featured GUT-level mass matrices (\ref{Mu})-(\ref{Mnu}).
\subsubsection{Integrating out the heavy degrees of freedom\label{sect:massmatrices}}
\vskip 2mm\paragraph{Up-type quarks:} Since there are no multiplets in the $10_{M}$, c.f. decompositions (\ref{decs}), with the up-type quark quantum numbers the effective MSSM up-quark mass matrix (evaluated at the GUT scale) is  identical to the $SO(10)$-level mass matrix (\ref{Mu}):
\bea
{\cal M}_u &=&  Yv^{10}_u \label{calMu}\,.
\eea
\vskip 2mm\paragraph{Down-type quarks and charged leptons:} The situation is very different though for down-type quarks and charged leptons whose GUT-level mass matrices (\ref{Md}) and (\ref{Me}) are $(3+n)\times (3+n)$-dimensional. They can be brought into a convenient form by means of  transformations   
\be
M_{d}\to M_{d}U_{d}^{\dagger}\equiv M_{d}', \;\;
M_{e}\to {U^{*}_{e}}M_{e}\equiv M_{e}'\label{Medprimed}, 
\ee
where $U_{d,e}$ are $(3+n)\times (3+n)$ unitary matrices such that $M'_{d}$ and $M'_{e}$ are block-triangular:
\be
M'_{d}={\cal O}\begin{pmatrix} v & v \\ 0 & M_{G} \end{pmatrix},\;\;
M'_{e}={\cal O}\begin{pmatrix} v & 0 \\ v & M_{G} \end{pmatrix}.\label{Mprimes}
\ee
This corresponds to the change of basis in the right-handed (RH) down quark and left-handed (LH) lepton sectors respectively: 
\bea
\begin{pmatrix}
d^c_L \\
\tilde{\Delta}^c_L
\end{pmatrix}
\equiv U_{d}
\begin{pmatrix}
D^c_L \\
\Delta^c_L
\end{pmatrix}
,\;\;
\begin{pmatrix}
\ell_L \\
\tilde{\Lambda}_L
\end{pmatrix}
\equiv U_{e}
\begin{pmatrix}
L_L \\
\Lambda_L
\end{pmatrix}.\label{Uedaction}
\eea
Here the upper components of the rotated vectors ($d^{c}_{L}$ and $\ell_{L}$) correspond to the light MSSM degrees of freedom. {Note also that the residual $SU(2)_{L}$ gauge symmetry makes the GUT-scale rotations (\ref{Uedaction}) act on both the charged lepton ($E_{L}$; $\Lambda_{L}^{-}$) as well as the neutrino ($N_{L}$; $\Lambda_{L}^{0}$) components of the leptonic doublets $L_{L}$ and $\Lambda_{L}$.}

Since the residual rotations acting on the LH quark and RH charged lepton components bringing the $M'_{d,e}$ matrices into fully block-diagonal forms are extremely tiny (of the $v/M_{G}$ order of magnitude) the $3\times 3$  upper-left blocks (ULB) in relations (\ref{Mprimes}) can be readily identified with the effective light down-type quark and charged lepton mass matrices, i.e., ${\cal M}_{d}\equiv (M_{d}')_{\rm ULB}$, ${\cal M}_{e}\equiv (M_{e}')_{\rm ULB}$. 
Given the specific form of $M_{d}$ and $M_{e}$ in eqs. (\ref{Md}) and (\ref{Me}) and parametrizing the unitary matrices $U_{d}$ and $U_{e}$ as 
\bea
U_{d,e}=\begin{pmatrix} A_{d,e} & B_{d,e} \\ C_{d,e} & D_{d,e} \end{pmatrix},\label{Ude}
\eea
(here $A_{d,e}$, $B_{d,e}$, $C_{d,e}$ and $D_{d,e}$ are $3\times 3$, $3\times n$, $n\times 3$ and $n\times n$ matrices, respectively) one obtains  
\begin{eqnarray}
{\cal M}_d &=& Y A^\dag_d v^{10}_d +  F B^\dag_dv_d^{16}, \label{calMd}\\
{\cal M}^T_e &=& Y A^\dag_e v^{10}_d +  F B^\dag_ev_d^{16}\label{calMe}. 
\end{eqnarray}
The off-diagonal GUT-scale blocks of $M_{d}$ and $M_{e}$ are rotated away provided
\begin{eqnarray}
F^T A^\dag_dV^{16}  + M_\Delta B^\dag_d & = & 0\,, \label{dZero}\\
F^T A^\dag_e V^{16} + M^T_\Lambda B^\dag_e & = & 0\,, \label{eZero}
\end{eqnarray}
which link the $A_{d,e}$ and $B_{d,e}$ factors.
The last two relations, together with the unitarity of $U_{d,e}$ implying
\be 
A_{d,e}A_{d,e}^{\dagger}+B_{d,e}B_{d,e}^{\dagger}=1,\label{unitarity}
\ee
impose strong constrains on the elements of matrices (\ref{Ude}) entering the effective mass formulas (\ref{calMd}) and (\ref{calMe}). These correlations shall be fully exploited in section \ref{sect:analysis}. 
\vskip 2mm\paragraph{Neutrinos:}
The situation in the neutrino sector is slightly more complicated due to the higher dimensionality of the GUT-level mass matrix (\ref{Mnu}). Notice, however, that the action of the LH leptonic rotation (\ref{Uedaction}), corresponding to a transformation $M_{\nu}\to U^*_\nu M_\nu U^\dag_\nu\equiv M'_{\nu}$ with $U_{\nu}$ denoting the relevant $(6+2n)\otimes(6+2n)$-dimensional unitary matrix, yields $M'_{\nu}$ in a hierarchical form\footnote{Note that the upper-right corner zero is due to the $SU(2)_{L}$ gauge symmetry which promotes the requirement (\ref{eZero}) of a similar zero in the charged lepton mass matrix (\ref{Medprimed})  to neutrinos.}
\bea
M'_\nu
=\begin{pmatrix}
 B^*_e \lambda B^\dag_e w_+ &   A^*_e Y  v^{10}_u &  B^*_e \lambda D^\dag_e  w_+ &  0 \\
\cdot  &  0  &  Y C^\dag_e v^{10}_u  &   F v_d^{16}\\
\cdot  &  \cdot  &  D^*_e \lambda D^\dag_e  w_+ & {M}_{\tilde{\Lambda}}\\
\cdot &  \cdot  &  \cdot  &   \lambda w_-
\end{pmatrix},\label{Mnuprimenaive}
\eea
with an abbreviation ${M}_{\tilde{\Lambda}}^{T}\equiv F^T C^\dag_e V^{16} + M^T_\Lambda D^\dag_e$ for the only GUT-scale entry therein. 

Na\"ively, given the hierarchies of the $SU(2)_{L}$ triplet, doublet and singlet VEVs, this shape of $M'_{\nu}$ yields three electroweak-scale pseudo-Dirac neutrinos at the effective theory level {(corresponding to the upper-left $6\times 6$ block of $M_{\nu}'$ above),} in an obvious conflict with observation. This is namely due to the fact that the lower-right $(3+2n)\times (3+2n)$-dimensional sector of  matrix (\ref{Mnuprimenaive}) corresponding to the $(N_{L}^{c},\tilde{\Lambda}^{0}_{L},\Lambda^{c0}_{L})$ part of the rotated basis does not have a full rank at the GUT scale. 

\subsubsection{Calculable triplet seesaw}
However, this issue should not be taken very seriously unless the quantum stability of the small entries in $M'_\nu$ is discussed. 
In particular, the 22 block zero (corresponding to the  $N_{L}^{c}N_{L}^{c}$ bilinear in $W_{Y}$, i.e., a SM singlet-singlet contraction) is not protected by the electroweak symmetry and thus can be naturally subject to large corrections which, eventually, may restore the full (GUT-scale) rank of the lower-right block of $M_{\nu}'$. 

For instance, a dimension 5 operator of the form {$16_{M}16_{M}\overline{16}_{H}\overline{16}_{H}/M_{P}$, where $M_{P}$ is the Planck} scale, lifts this zero sufficiently to change the entire picture: the lower-right block becomes superheavy and the hierarchical matrix structure \`a la standard seesaw is achieved. Subsequently, one is left with three sub-eV Majorana neutrinos at the SM level, with the upper-left entry of $M_{\nu}'$ promoted to the role of an additive (type-II-like) contribution to their effective mass matrix.   

Let us also remark that a simple renormalizable realization of this scheme is obtained if the matter sector is further extended by three $SO(10)$ singlets, well in the spirit of $E_{6}$ gauge models. The extra contraction ${16}_M  1_{M}\overline{16}_H$ in the Yukawa superpotential   
provides the necessary set of large matrix elements entering the heavy part of the (extended) neutrino mass matrix even at the renormalizable level. 

However, given the likely proximity of such a new physics scale to $M_{G}$, one expects other physical effects to affect all the effective mass matrices at some level. 
{Obviously, it is not very appealing to let the non-renormalizable operators and/or similar effects into play in the simple scheme of our interest unless these are under a very good control\footnote{{Note that giving up renormalizability one would actually loose a great deal of the original motivation for the vector-like matter entering the genesis of the SM flavour structure, as discussed in section \ref{sect:introduction}. Indeed, there is a lot of non-renormalizable $SO(10)$ models of flavour in the literature with spinorial matter only.}}. Actually, as we have already emphasized, the goals of the current analysis are rather different and, as long as we focus on the renormalizable part of the effective flavour structure, a deep understanding of all the neutrino sector details is not strictly required.
} 

Indeed, whatever the ultimate rank-restoration mechanism happens to be, the seesaw contribution due to the $SU(2)_{L}$ triplet in $54_{H}$, 
\be
{\cal M}^{\Delta}_\nu \equiv B^*_e \lambda B^\dag_e w_+\label{calMnuII}\,,
\ee
is always present and the underlying $10_{M}10_{M}54_{H}$ contraction is particularly robust. Indeed, apart from the standard $SU(2)_{L}\otimes U(1)_{Y}$ gauge symmetry protection, this is namely due to the fact that the triplet VEV within $54_{H}$ can not be mimicked by $\langle 45^{2}_{H}\rangle$ nor $\langle\overline{16}_{H}^{2}\rangle$ at the $d=5$ level. 
Since ${\cal M}^{\Delta}_\nu$ is also the only calculable part of the effective neutrino mass matrix in the simple framework of our interest, the best one can do is to focus entirely on it and assume its dominance over the other contributions in ${\cal M}_\nu$:
\bea
{\cal M}_\nu\sim {\cal M}^{\Delta}_\nu\,.
\eea
This approximation is what we shall adopt from now on. Let us also note that a dedicated analysis of the conditions under which such situation can be realized in a specific complete model is a highly non-trivial enterprise, much beyond the scope of this work.  
\section{Analysis and discussion\label{sect:analysis}}
\subsection{General prerequisites and comments\label{sect:generalities}}
With all this information at hand one can attempt to exploit the strong correlations between the effective mass matrices (\ref{calMu}), (\ref{calMd}), (\ref{calMe}) and (\ref{calMnuII}) to assess the viability of the general framework by means of a global $\chi^{2}$ analysis of its compatibility with the measured quark and lepton masses and mixings. 
\subsubsection{The effective quark and lepton mass matrices}
Before that, one should attempt to further simplify the relevant mass matrices (\ref{calMd})-(\ref{calMe}). First, one can substitute $Y$ for ${\cal M}_{u}$ and eliminate the $B_{d,e}$ factors in eqs. (\ref{calMd}), (\ref{calMe}) and (\ref{calMnuII}) by using relations (\ref{dZero}) and (\ref{eZero}) so that $A_{d,e}$'s remain the only ``complicated'' factors in all formulas of our interest:
\bea
\tilde{\cal M}_d &=& \left( r \tilde{\cal M}_u - \tilde{F} (\tilde{M}_\Delta)^{-1} \tilde{F}^T \right) A^\dag_d \label{calMd2} , \\
\tilde{\cal M}^T_e &=& \left( r \tilde{\cal M}_u - \tilde{F} (\tilde{M}^T_\Lambda)^{-1} \tilde{F}^T \right) A^\dag_e \label{calMe2}
,\\
\tilde{\cal M}_\nu &\propto &A^*_e \tilde F (\tilde{M}_\Lambda )^{-1} \tilde{\lambda} (\tilde{M}^T_\Lambda )^{-1} \tilde{F}^T A^\dag_e, \label{calMnu2}
\eea
with $r\equiv v_{d}^{10}/v_{u}^{10}$, $\tilde{\cal M}_{u,d,e}\equiv {\cal M}_{u,d,e}/{m}_{b}$, $\tilde{F}\equiv F v_d^{16}/{m}_{b}$, $\tilde{M}_{\Delta,\Lambda}\equiv (v_d^{16}/{m}_{b}){M}_{\Delta,\Lambda}/V^{16} $ and $\tilde{\lambda}\equiv (v_d^{16}/{m}_{b})(V^{54}/V^{16})\lambda$ where ${m}_{b}$ stands for the bottom quark mass. In what follows, it will also be convenient to normalize the antisymmetric parts of $\tilde{M}_{\Delta,\Lambda}$ in the same manner:
{$\tilde{\eta}_{\Delta,\Lambda}\equiv (v_d^{16}/{m}_{b})(V^{45}_{\Delta,\Lambda}/V^{16})\eta$}.
Note that the overall scale of $\tilde{\cal M}_{\nu}$ driven by $w_{+}$ remains undetermined at the current level. For this reason we have dropped the explicit triplet VEV and introduced a proportionality sign into eq. (\ref{calMnu2}). 

It should be also possible to write down the $A_{d,e}$ factors in terms of the {Yukawa} superpotential parameters as we did for the brackets in eqs. (\ref{calMd2}) and (\ref{calMe2}) which, however, could be quite complicated in general . Actually, we don't need to do so as there are redundancies in $A_{d,e}$ that do not play any role in the low energy phenomenology (i.e., spectra and LH mixings). Indeed, one can always decompose $A_{d,e}$ as 
\be
A_{d,e}=V_{d,e}H_{d,e}\label{Adedecomposition}
\ee
where $V_{d,e}$ and $H_{d,e}$ are unique $3\times 3$ unitary and hermitean matrices respectively.
It is clear that $V_{d,e}$ do not affect the low energy quark and lepton observables because $V_{d}$ contributes only to the RH quark rotations and $V_{e}$ enters $\tilde{\cal M}_{e}$ and $\tilde{\cal{M}}_{\nu}$ on the same footing and thus cancels in the leptonic mixing matrix. 

Given (\ref{Adedecomposition}) the hermitean factors $H_{d,e}$ can be determined from the unitarity of  $U_{d}$ and $U_{e}$ (\ref{unitarity}) taking into account the triangularization constraints (\ref{dZero}) and (\ref{eZero}):\footnote{The square root of a generic hermitian positive semidefinite matrix $M$ is defined as $U\sqrt{D}U^{\dagger}$ where $D=U^{\dagger}MU$ is a real non-negative diagonal matrix. Note that the sign ambiguity in $\sqrt{D}$ does not play any role due to the irrelevance of the overall signs of the generalized eigenvalues of matrices (\ref{calMd2})-(\ref{calMnu2}) and the corresponding mixing angles.} 
\bea
H_{d}&=&(1+\tilde{F}^{*}(\tilde{M}_{\Delta} \tilde{M}_{\Delta}^{\dagger})^{-1}\tilde{F}^{T})^{-1/2}\label{Hd}\,,\\
H_{e}&=&(1+\tilde{F}^{*}(\tilde{M}_{\Lambda}^{T} \tilde{M}_{\Lambda}^{*})^{-1}\tilde{F}^{T})^{-1/2}\label{He}\,.
\eea
To conclude, formulas (\ref{calMd2})-(\ref{He}) admit for a full reconstruction of the quark and lepton masses and mixing (up to the absolute neutrino mass scale and irrelevant basis transformations $V_{d,e}$) for any point in the parametric space of the model. 
\subsubsection{Basic features and strategy for potentially realistic fits\label{structure}}
Let us now comment on the salient features of the effective flavour structure (\ref{calMd2})-(\ref{calMnu2}) and its prospects for accommodating successfully the quark and lepton data. 
\begin{itemize}
\item First, it is clear that for non-zero $\tilde{F}$ and $\tilde{M}_{\Delta,\Lambda}$ the up and down quark mass matrices as well as the hierarchies of their spectra are different and a non-trivial quark mixing is generated. 
\item The Cabibbo-Kobayashi-Maskawa (CKM) quark mixing angles are naturally generated when the magnitude of the $\tilde{F} (\tilde{M}_\Delta)^{-1} \tilde{F}^T$ term in (\ref{calMd2}) is smaller than that of $r \tilde{\cal M}_{u}$; otherwise the approximate alignment of $\tilde{\cal M}_{u}$ and $\tilde{\cal M}_{d}$ is lost and there is no reason for the CKM mixing to be small. 
\item In such settings, the $r$ parameter has a clear interpretation of a ``hierarchy compensator'' between $m_{b}$ and $m_{t}$ and as such its value is strongly constrained. As we shall recapitulate  in section~\ref{sect:nogo} (c.f. \cite{Malinsky:2008zz}), pushing $r$ out of this natural domain hampers the prospects of getting good fits of both the down quarks and the charged leptons at once.
 
\item The case of  sub-leading $\tilde{F} (\tilde{M}_\Delta)^{-1} \tilde{F}^T$ and $\tilde{F} (\tilde{M}_\Lambda^{T})^{-1} \tilde{F}^T$ naturally accommodates the 
approximate convergence of the $b$ and $\tau$ Yukawa couplings observed in many studies of the running of Yukawa couplings. 
\item
Moreover, for $\tilde F$  in the ${\cal O}(1)$ ballpark, the same implies $\tilde{F}^{*}(\tilde{M}_{\Delta} \tilde{M}_{\Delta}^{\dagger})^{-1}\tilde{F}^{T},\tilde{F}^{*}(\tilde{M}_{\Lambda}^{T} \tilde{M}_{\Lambda}^{*})^{-1}\tilde{F}^{T}\ll 1$, which provides a further insight into the effective mass formulas for $\tilde{\cal M}_{d}$ and $\tilde{\cal M}_{e}$ because it renders the $A_{d}^{\dagger}$ and $A_{e}^{\dagger}$ factors in (\ref{calMd2})-(\ref{calMe2}) unimportant even for the second generation.
\item With $\tilde{F} (\tilde{M}_\Delta)^{-1} \tilde{F}^T$ and $\tilde{F} (\tilde{M}_\Lambda^{T})^{-1} \tilde{F}^T$ in a few percent domain there should be enough room to accommodate the differences among $m_{c}/m_{t}$, $m_{s}/m_{b}$ and $m_{\mu}/m_{\tau}$. Moreover, even the basic hierarchy between the CKM mixing angles $\theta_{12}\gg \theta_{23,13}$ seems very natural: with a diagonal $\tilde{\cal M}_{u}$ the only CKM angle that can be large due to a few-percent off-diagonalities from the sub-leading term is $\theta_{12}$. 
\item The neutrino mass matrix (\ref{calMnu2}) has nothing to do with the leading contribution to $\tilde{\cal M}_{e}$ and thus there is no reason for the leptonic mixings to be small.
\end{itemize}
Remarkably, this scheme matches perfectly the basic qualitative features of the observed quark and lepton mass and mixing pattern. In what follows, we shall be using the values given in TABLE \ref{tab:inputs} as physical inputs of the numerical analysis carried out in section~\ref{sect:numerics}.
\renewcommand{\arraystretch}{1.2}
\begin{table}[th]
	\centering
	\scalebox{1.00}{
  \begin{tabular}{|c|c|c|c|}
  \hline\hline
  \multicolumn{4}{|c|}{Quark sector}\\
  \hline
 observable & value & observable & value \\
  \hline
  $m_u$ [MeV]   					& $0.45 (\pm 0.2)$ 		& $m_d$ [MeV]   						& $1.3 \pm 0.6$\\
  $m_c$ [MeV]   					& $217 (\pm 35)$ 			& $m_s$ [MeV]   						& $23 \pm 6$\\
  $m_t$ [GeV]   					& $97 (\pm 38)$ 			& $m_b$ [GeV]   						& $1.4\pm 0.6$ \\
  $\sin\theta^{q}_{12}$ 	& $0.2243\pm 0.0016$ 	& $\sin\theta^{q}_{23}$ 		& $0.0351\pm 0.0013$\\
  $\sin\theta^{q}_{13}$ 	& $0.0032\pm 0.0005$ 	& $\delta_\mathrm{CP}^{q}$  & $60^\circ \pm 14^\circ$\\
 \hline
  \multicolumn{4}{|c|}{Lepton sector}\\
  \hline
 observable & value & observable & value \\
  \hline
  $\Delta m^2_{21}$ [eV$^2$]    & $(7.7 \pm 0.2)\, 10^{-5}$ 	& $m_{e}$ [MeV] 					& 0$.3565\pm 0.0100$ \\
  $|\Delta m^2_{31}|$ [eV$^2$] 	& $(2.40\pm 0.12) 10^{-3}$  	& $m_{\mu}$ [MeV] 				& $75.3\pm 1.2$\\
 	$\sin^2\theta^{l}_{12}$ 			& $0.304\pm 0.019$ 						& $m_{\tau}$ [GeV] 				& $1.629\pm 0.037$ \\
	$\sin^2\theta^{l}_{23}$ 			& $0.50\pm 0.06$  						& $\sin\theta_{13}^{l}$ 	& {$\leq 0.18$}\\ 
 \hline\hline
  \end{tabular}
		}
	\caption{Sample GUT-scale inputs of the numerical \hyphenation{ana-lysis}analysis performed in section\ref{sect:numerics}. The specific values correspond to those given in \cite{Malinsky:2005zz} for the quark sector and \cite{running} 
(c.f. also \cite{Xing:2007fb}) for the charged lepton masses, $\tan\beta=55^{\circ}$.  The solar and atmospheric neutrino mass squared differences and the leptonic mixings are taken from \cite{Schwetz:2008er}. {The upper bound on $\theta_{13}^{l}$ corresponds to the global $90\%$ C.L. value quoted in \cite{theta13}. }The running effects in the neutrino sector have been neglected due to the hierarchical shape of the neutrino spectrum. For sake of simplicity,  symmetric $\sigma$-ranges have been adopted. The error in the electron mass has been artificially enhanced by a factor of 10 to improve the convergence of the numerics, with no significant impact on the quality of the actual fits. 
	}
	\label{tab:inputs}
\end{table}
\subsubsection{Parameter counting\label{sect:parametercounting}}
In order to assess the prospects of testing this picture even at the quantitative level it is worth  counting the number of independent parameters. Working with a real and positive $\tilde{\cal M}_{u}$  (that fixes entirely the basis in the space of $SO(10)$ matter spinors) the phase of $r$ can be rotated away by a global phase redefinition of $\tilde{\cal M}_{d}$ and $\tilde{\cal M}_{e}$, leaving a single real parameter (RP). A similar rotation in the space of $n$ $SO(10)$ matter vectors can bring the $M_{10}$ matrix to the real and diagonal form with $n$ RPs. In this basis, the complex symmetric Yukawa coupling of $54_{H}$ ($\lambda$) adds $(n+1)n$ RPs and the antisymmetric Yukawa of $45_{H}$ ($\eta$), which is present for $n\geq 2$, yields $(n-1)n$ RPs. For  $n\geq 2$, one must also add the complex ratio of the two VEVs in $45_{H}$, accounting for an extra pair of RPs. Finally, there is the $3\times n$-dimensional complex matrix of $\tilde F$'s adding in general $6n$ RPs. In total, one ends up with $2n^{2}+7n+3$ RPs for $n\geq 2$ (and 10 RPs for $n=1$, in agreement with \cite{Malinsky:2008zz}). 

With the up-quark masses as inputs, there are 13 low energy observables one can attempt to fit (3 down-quark masses plus 4 CKM parameters in the quark sector, 3 charged lepton masses, the $\Delta m^{2}_{21}/|\Delta m^{2}_{31}|$ ratio in the neutrino sector and {2} leptonic mixing angles measured so far). A successful fit of these data could then admit to tell something about the unknown parameters (in particular, $\sin\theta_{13}^{l}$ {and the leptonic CP phases}).    
\subsection{Single active vector matter multiplet}
Let us begin with the case of a single vector matter multiplet in the game\footnote{
Since this case has been analysed in detail in \cite{Malinsky:2008zz} here we shall just recapitulate the salient features of this basic setting.}. 
For $n=1$, however, the neutrino mass matrix (\ref{calMnu2}) has rank 1 and thus there is no point in attempting to fit $\Delta m^{2}_{21}/|\Delta m^{2}_{31}|$ nor $\theta_{12}^{l}$. Hence, in full generality, one is left with 10 parameters to fit 11 observables, which clearly indicates a potential difficulty with a full-fledged three-generation fit. Nevertheless, since in practice there can easily be other $10_{M}$'s around (though perhaps at the verge of decoupling) it still makes sense to look at the two heavy generations.  As we shall see in section~\ref{sect:2by2}, an interesting link between the maximality of the atmospheric mixing in the lepton sector and the interplay among the 23 mixing in the quark sector and the $m_{s}/m_{b}$ ratio can emerge even in this obviously oversimplified case. Moreover, in order to appreciate the naturalness of the $n=2$ fits discussed in section~\ref{sect:numerics}, it is instructive to see explicitly where the trouble with the three-generation fit for $n=1$ \cite{Malinsky:2008zz} comes from; an analytic argument will be given in section~\ref{sect:nogo}.

\subsubsection{Triplet seesaw and a large 2-3 mixing in the $2\times 2$ case \label{sect:2by2}}
Perhaps the most intriguing feature of the minimal scenario is the simple correlation between the large values of the leptonic 2-3 mixing inherent to the triplet-dominated neutrino masses and the specific flavour structure observed in the 2-3 part of the quark sector. It reads
\be\label{atm}
\tan 2\theta_{23}^{l}\approx 2 |x| \big/ \left|1-x^{2}\right| \,,
\ee
with $x\equiv ({y_{b}}/{y_{s}})\sin\theta_{23}^{q}$, where $y_{s,b}$ are the Yukawa couplings of the heavy down-type quarks (in the diagonal basis) and $\sin\theta_{23}^{q}$ is the 2-3 mixing angle in the quark sector.
In a certain sense, this relation can be viewed as a ``radiatively stable'' analogue of the well-known Bajc-Senjanovic-Vissani (BSV) relation $\tan 2\theta_{23}^{l}\approx \sin 2\theta^{q}_{23} \big/2\sin^{2}\theta^{q}_{23}+\epsilon$ (with $\epsilon\equiv 1-y_{\tau}/y_{b}$) \cite{BSV} 
derived in the minimal SUSY $SO(10)$ GUT framework\footnote{What we mean here by ``radiative stability'' is that the  $y_{b}/y_{s}$ ratio is subject to a much milder running than the ratio $y_{\tau}/y_{b}$ underpinning the BSV relation in the minimal SUSY SO(10).}.
The relation of our interest (\ref{atm}) is readily obtained from the basic formulas for the charged lepton and the triplet neutrino masses (\ref{calMe2}) and (\ref{calMnu2}) taking into account the estimated structure of the charged sector fits specified in section~\ref{structure} or in~\cite{Malinsky:2008zz}. For sake of simplicity, we shall  also assume a CP-conserving setting with all phases either $0$ or $\pi$. At the leading order, the flavour structure of the triplet-dominated neutrino mass matrix can be approximated by
\be
\tilde{\cal M}_{\nu}\propto B_{e}^{*}\tilde\lambda B_{e}^{\dagger}\propto A^{*}_{e}\tilde F\tilde F^{T}A^{\dagger}_{e}\approx V^{*}_{e}\tilde F\tilde F^{T}V^{\dagger}_{e}\,,
\ee
 where we made use of the fact that $\tilde \lambda$ is a number now and  the ``external'' factors $A_{e}$ are almost unitary, see section~\ref{structure}.
Rotating away the $V_{e}$ matrices, the charged lepton mass matrix (\ref{calMe2}) becomes close to diagonal. Thus, focusing entirely on the 2-3 mixing (which, indeed, is the only leptonic angle it makes sense to look at with a rank=1 mass matrix),  it is almost entirely encoded in the neutrino mass matrix $\tilde{\cal M}_{\nu}\propto \tilde F \tilde F^{T}$ and one can write
\be\label{atmraw}
\tan 2\theta_{23}^{l}\approx 2 |\tilde{F}_{2}\tilde{F}_{3}| \big/ |\tilde{F}_{2}^{2}-\tilde{F}_{3}^{2}|\,.
\ee
In order to get a grip on the typical values of the $F$-parameters in (\ref{atmraw}) one should exploit the quark sector sum-rule. At the same level of accuracy as before, the relevant formula (\ref{calMd2}), once contracted to the 2nd and 3rd generations, yields
\be\label{MDapprox}
\tilde{\cal M}_{d}\approx r\left(
   \begin{array}{cc} 
      m_{c}/m_{b} & 0 \\
        0 & m_{t}/m_{b}
   \end{array}
\right)+\rho \left(
   \begin{array}{cc} 
      \tilde{F}_{2}^{2} & \tilde{F}_{2}\tilde{F}_{3}  \\
        \tilde{F}_{2}\tilde{F}_{3} & \tilde{F}_{3}^{2} 
   \end{array}
\right),
\ee
where $\rho\equiv (m_{b}/v_d^{16})V^{16}/M_{\Delta}$ and the approximate diagonality of $\tilde{\cal M}_{d}$ in the $\tilde{\cal M}_{u}$-diagonal basis has been used. Due to the estimated smallness of $r$ (in the few $\%$ range), one can expect (c.f. section~\ref{structure}) that the only relevant entry of the first matrix in (\ref{MDapprox}) is $m_{t}/m_{b}$. The resulting $\tilde{\cal M}_{d}$ can be easily shown to give $m_{s}/m_{b}\approx \rho \tilde{F}_{2}^{2}$, $1\approx r m_{t}/m_{b}$ and $\sin\theta_{23}^{q}\approx \rho \tilde{F}_{2}\tilde{F}_{3}$. Solving for $\tilde{F}_{2}$ and $\tilde{F}_{3}$ and substituting into (\ref{atmraw}) one recovers (\ref{atm}).
\subsubsection{The renormalizable $3\times 3$ charged sector no-go\label{sect:nogo}}
With such an observation at hand one would naturally ask whether the analysis can be extended to the  $3 \times 3$ case so that it might account for the details associated to the light flavours. 
Unfortunately, the answer is negative.  
The reason is that with a single vector matter multiplet at play there  is a fundamental obstacle to any potentially successful fit already at the charged sector level. Remarkably enough, one can even provide a simple analytic argument for why this happens to be so. 

For the sake of that, let us look at the shape of the down-quark mass matrix (\ref{calMd2}) and consider the three main minors of $\tilde{\cal M}_{d}\tilde{\cal M}_{d}^{\dagger}$ defined as $\Delta_{ij,i<j}\equiv d_{ii}d_{jj}-d_{ij}d_{ji}=d_{ii}d_{jj}-|d_{ij}|^{2}$ with $d_{ij}\equiv (\tilde{\cal M}_{d}\tilde{\cal M}_{d}^{\dagger})_{ij}$.  Notice that these quantities, by definition, depend only on the physical inputs, in particular the quark masses and mixing parameters. One can easily show that the $\tilde{F}$ couplings enter  $\Delta_{ij,i<j}$ only as $|\tilde{F}_{i}|^{2}|\tilde{F}_{j}|^{2}$ (recall there is only a single $10_{M}$ here so $\tilde{F}$ is a vector) and one can solve the three relations for $\Delta_{ij,i<i}$ for these factors:
\be\label{bijs}
\rho^{2}|\tilde{F}_{i}|^{2}|\tilde{F}_{j}|^{2}=\frac{|d_{ij}|^{2}-\left(d_{ii}-r^{2}m_{u}^{i2}\right)\left(d_{jj}-r^{2}m_{u}^{j2}\right)}{r^{2}\left[m_{u}^{i2}+m_{u}^{j2}-2m_{u}^{i}m_{u}^{j}\cos (\gamma_{i}-\gamma_{j})\right]}\,,
\ee
where $\gamma_{i}$ are the phases of  $\tilde{F}_{i}$ defined as $\tilde{F}_{i}\equiv|\tilde{F}_{i}|e^{-i\gamma_{i}/2}$. It is clear that consistency requires the numerators on the RHS of eq.~(\ref{bijs}) to be non-negative for all $i,j$. First, this can never be realised nontrivially if the CKM mixing was turned off (implying $d_{ij,i\neq j}=0$) -- at least one pair out of  any three non-zero numbers always yields a positive product. Thus, {\it with a single $10_{M}$ at hand, a non-trivial $V_{CKM}$ is a necessary condition for any successful quark sector fit}. Second, turning on the small CKM mixing, the numerators look like products of pairs of quadratic functions in $r^{2}$ with small positive shifts due to $|d_{ij}|^{2}\neq 0$. Taking into account the physical ranges of the quark masses and mixing angles it is straightforward to check that the only domain, in which all three of these expressions can be simultaneously positive, corresponds to $r\approx m_{s}/m_{c}\sim 0.15$. This value, however, is one order of magnitude away from the physically motivated expectation identified in section~\ref{structure}, at odds with the desired shape of the  charged lepton spectrum. Thus, even at the pure charged sector level, there is a generic no-go for the fits of the flavour structure of the minimal model with a single vector matter multiplet \cite{Malinsky:2008zz}. 

On the practical side, one should emphasize that the argument above is based on the specific values of the input parameters used throughout this analysis, c.f. TABLE \ref{tab:inputs}. These, however, depend on a particular scenario employed to study their running properties. For instance, large SUSY thresholds \cite{largeSUSYeffects} can significantly affect the GUT-scale mass ratios, especially for the light generations, and, hence, the desired range for the $r$ parameter. Thus, at least in principle, there could still be an option for the $n=1$ case to be implemented in models yielding unconventional high-scale Yukawa patterns. A detailed discussion of these issues, however, is beyond the scope of this work.

\subsection{Two active vector matter multiplets\label{sect:numerics}}
In view of the negative result for $n=1$ it is natural to ask whether the charged-sector no-go can be overcome with more than a single extra matter multiplet in the game, in particular with $n=2$,  and if yes how much one can learn about the leptonic mixing (especially about $\theta_{13}^{l}$) {and CP violation} in such case.  
At first glance, one would expect the $n=2$ fits to be essentially trivial as the dimensionality of the parametric space increases dramatically: from 10 for $n=1$ to 25 for $n=2$, c.f. section \ref{sect:parametercounting}.
On the other hand, the 3 extra constraints from the leptons can play an important role, given the qualitative difference among the hierarchies and mixings in the quark and lepton sectors.  Moreover, as we know from the previous section, good fits are impossible if the second $10_{M}$ plays only a marginal role, i.e., if it dynamically decouples. 

To put this statement on a firm ground one should take into account how the gauge-singlet mass parameters encoded in the $M_{10}$ matrix enter the heavy matter spectrum. In the normalization $\tilde{M}_{10}\equiv (v_d^{16}/{m}_{b}){M}_{10}/V^{16}$ (see section\ref{sect:generalities}) one can conveniently parametrize 
\be\label{tildeM10}
\tilde M_{10}=t\,{\rm diag(1,p)}\;,
\ee 
where $t$ is an overall factor and $p$ encodes the hierarchy of the two eigenvalues of $\tilde{M}_{10}$. Note that one can take $p\geq1$ without loss of generality. Then, $p\to \infty$ corresponds to the decoupling limit if the couplings between the heavy and the light GUT-scale matter states are kept under control, i.e., do not diverge.  
The expected worsening of the best $\chi^{2}$ towards the decoupling limit can then be used as a non-trivial consistency check of the numerical results we shall present in the subsequent sections.

\subsubsection{Fits with $45_{H}$ decoupled from the Yukawa sector\label{sect:54only}}
Let us begin with the case of  a negligible contribution from the $45_{H}$ Yukawa coupling. As we shall see in section~\ref{sect:54plus45}, this is well motivated because of a high degree of ``sterility'' of $45_{H}$ in the $n=2$ fits {whenever there is more than an ${\cal O}(1)$ hierarchy between the mass terms of the two $10_{M}$.} 
Note also that the situation with {$\eta\to 0$} is effectively parametrized by only 21 RPs; thus, taking into account the strong phenomenology constraints on $r$ together with the perturbativity bounds on $|F_{ik}|$'s and the limited impact of their phases, it is actually far from clear whether this setting admits good fits.
In what follows, we shall use a simple prescription for the relevant decoupling parameter\footnote{The specific form of the definition (\ref{decoupling}) corresponds to the role the $c_{22}$ and $c_{12}$ factors play in the spectrum of $\tilde{M}_{\Delta,\Lambda}$ which (for $c_{11}$ in the ${\cal O}(1)$ domain) is well approximated by their determinants (linear in $p$ and $c_{22}$ and quadratic in $c_{12}$).}, 
\be
P=p/{\rm max}\{|c_{22}|,|c_{12}|^{2}\} \label{decoupling}\,,
\ee
where $c_{kl}$ govern the entries of the properly normalized Yukawa coupling of $54_{H}$,
\be
\tilde {\lambda}_{kl}\equiv t\, c_{kl}\,.
\ee  
The shape of formula (\ref{decoupling}) reflects the basic features of the numerical fits, namely the dominance of $|c_{22}|\propto p$ for large $p$ followed by a milder behaviour of $|c_{12}|\propto \sqrt{p}$ and an essentially $p$-insensitive $|c_{11}|\propto p^{0}$. Apart from the $c_{12}$ playing the obvious ``destructive'' role of mixing up the heavy and the light sectors, the $c_{22}$ is taken into account because it can mimic {an ``effective'' $p$ in $\tilde{M}_{\Delta}$ or $\tilde{M}_{\Lambda}$.}   
\vskip 2mm\paragraph{Pure charged sector fits - avoiding the $n=1$ no-go: \label{sect:n2:54onlychargedleptons}}
\begin{figure}[th]
\includegraphics[width=8.5cm]{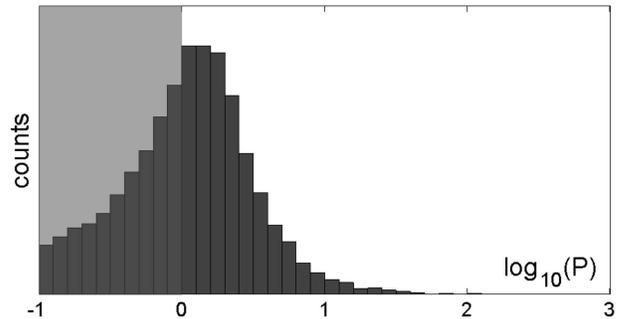}
\caption{\label{histogram}An  $n=2$ histogram of the relative frequency of fits obtained  for the quark and charged lepton masses and CKM mixing parameter with $\chi^{2}<1$ for different values of the decoupling parameter $P$. The sharp decline of the counts towards the high $P$ limit is a manifestation of the no-go discussed in section {\ref{sect:nogo}} for the  $n=1$ case.  {The shaded region on the left corresponds to the fine-tuned setting with $V^{54}$ dominating over the singlet mass parameter $M_{10}$ and its specific shape is an artefact of the numerical method we use.}}
\end{figure}

The first test to be passed concerns the charged lepton fits in the $n=2$ case. Recall that in section~\ref{sect:nogo} these were shown to be generally troublesome in the $n=1$ case despite the relatively large number of parameters (10 in general) available to fit just 7 observables (3 down-type quark masses and 4 CKM mixing parameters if, for the sake of simplicity, the up-type quark masses are fixed at their means). With the extra $10_{M}$ at hand, excellent fits are easily obtained within the expected domains (see section~\ref{structure}) whenever its contribution is non-negligible. Quantitatively, as seen in FIG.~\ref{histogram}, we have found good fits of the charged sector data for all values of the decouplings parameter $P$ below about one hundred. In other words, the value of {$P^{-1}\sim 1\%$} constitutes a qualitative boundary above which the second vector matter multiplet is already decoupled too much to avoid the no-go inherent to the $n=1$ settings.

{In this respect, it is also worth noting that the interesting link between the large atmospheric mixing and the specific value of $m_{b}/m_{s}\sin\theta_{23}^{q}\approx 1$ obtained in the $n=1$ case, c.f. section~\ref{sect:2by2}, is upset due to the perturbations coming from the second vector matter multiplet and there is no preferred value of $\theta_{23}^{l}$ observed in these fits.}

\vskip 2mm\paragraph{\!\!\! Fits including leptonic $\theta_{12}^{l}$, $\theta_{23}^{l}$ and $\Delta m^{2}_{21}/|\Delta m^{2}_{31}|$: \label{sect:n2:54only}}
{Including from now on the measured values of the relevant neutrino oscillation parameters, i.e., $\theta_{12}^{l}$, $\theta_{23}^{l}$ and $\Delta m^{2}_{21}/|\Delta m^{2}_{31}|$, into the $\chi^{2}$ function one can still attempt to get predictions for $\theta_{13}^{l}$ and the leptonic Dirac ($\delta_{CP}^{l}$) and Majorana CP phases.}
{Remarkably enough, such fits} turn out to be nontrivial in spite of the high number of free parameters at play. This is reflected by the fact that none of the fits we obtained yields $\chi^{2}$ below around {15; nevertheless, given the number of fitted observables, these values are still to be regarded as very good.}  

\begin{figure}[h]
\includegraphics[width=8.6cm]{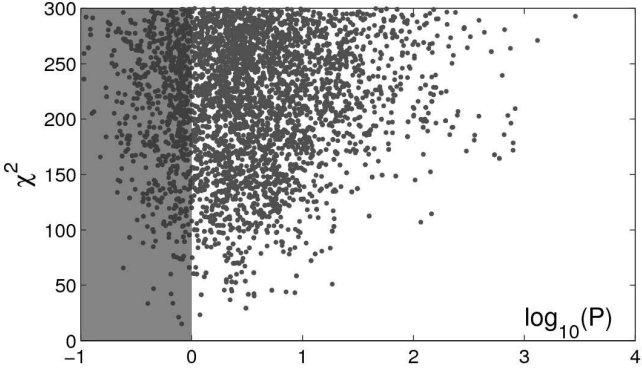}
\caption{{A sample of the $\chi^{2}$ values for the $n=2$ fits with
{$\eta\to 0$} as a function of the decoupling parameter $P$. One can see clearly that the extra constraints from the leptonic sector make the $n=2$ fits non-trivial, see also section~\ref{sect:54plus45} and TABLE~\ref{tab:BestSol_54only}.}\label{fig:chisqxP}}
\end{figure}

The generic behaviour of the relevant fits can be seen in FIG.~\ref{fig:chisqxP}. Now, the best $\chi^{2}$ value is a steeply rising function of the decoupling parameter $P$, in agreement with expectation. On the more technical side, here we have also decided to lift some of the residual degeneracies in the parametric space by further constraining the $c_{kl}$ parameters into the ${\cal O}(1)$ domain, which provides a convenient link between the $p$ and $P$ parameters. A detailed information about a pair of the relevant best-$\chi^{2}$ solutions is given in TABLE~\ref{tab:BestSol_54only}. 

\begin{table}[htbp]
	\centering
	\scalebox{1}{
{
		\begin{tabular}{|c|c|c|}
			\hline
			Parameter & Fit I & Fit II \\
			\hline
			\hline
			$p$	& $14.339010$ & $2.847552$ \\ 
			\hline\hline 
			$r$ & $0.01621150$ & $0.01473598$ \\ 
			$t$ & $31.791794$ & $162.846941$ \\ 
			\hline
			$\tilde{m}_u$ & $0.0003270355$ & $0.0003340457$ \\
			$\tilde{m}_c$ & $0.1618762$ & $0.1698742$ \\
			$\tilde{m}_t$ & $69.008012$ & $78.470173$ \\ 
			\hline
			$\tilde{F}_{11}$ & $-0.169686 - 0.163037i$ & $-0.398016 - 0.499151i$ \\
			$\tilde{F}_{21}$ & $ 0.568262 + 1.543220i$ & $-0.322017 -     0.584528i$ \\
			$\tilde{F}_{31}$ & $-2.992396 + 1.508342i$ & $ 4.433113 - 1.105430i$ \\ 		
			$\tilde{F}_{12}$ & $ 0.300966 + 0.871563i$ & $ 0.997104 - 0.219207i$ \\ 
			$\tilde{F}_{22}$ & $-1.198678 - 2.197259i$ & $ 3.814132 +     0.370594i$ \\ 
			$\tilde{F}_{32}$ & $-1.731386 - 0.158647i$ & $ 7.384473 -      3.331683i$ \\ 			
			\hline
			$c_{11}$ & $ 1.958581 - 3.831928i$ & $-0.766690 - 1.509670i$ \\
			$c_{22}$ & $-0.550307 - 0.871218i$ & $-0.031916 + 0.033495i$ \\
			$c_{12}$ & $ 2.555502 - 3.497512i$ & $ 0.769329 - 1.704622i$ \\
			\hline
			\hline
			$m_u$ [MeV]   						& $0.4579$ 			& $0.4677$\\
  		$m_c$ [MeV]   						& $226.6$ 			& $237.8$\\
  		$m_t$ [GeV]   						& $96.61$ 			& $109.86$\\
  		\hline
			$m_d$ [MeV]   						& $0.8892$ 			& $0.9909$\\
  		$m_s$ [MeV]   						& $40.24$ 			& $30.50$\\
  		$m_b$ [GeV]   						& $1.461$ 			& $1.634$\\
  		\hline
   		$\sin\theta^{q}_{12}$ 		& $0.2248$ 			& $0.2240$\\
   		$\sin\theta^{q}_{23}$ 		& $0.03487$ 		& $0.03153$\\
  		$\sin\theta^{q}_{13}$ 		& $0.003304$ 		& $0.003958$\\
  		$\delta_\mathrm{CP}^{q}$ 	& $37.38^\circ$ & $60.83^\circ$\\
			\hline
			\hline
			$m_{e}$ [MeV] 						& $0.3561$ 			& $0.3582$\\
  		$m_{\mu}$ [MeV] 					& $75.29$ 			& $75.32$ \\
 			$m_{\tau}$ [GeV] 					& $1.630$ 			& $1.588$\\
 			\hline
  		$\frac{\Delta m^2_{21}}{|\Delta m^2_{31}|}$ & $0.03269$  & $0.03244$\\
  		\hline
 			$\sin^2\theta^{l}_{12}$ 	& $0.2714$ 			& $0.3031$\\
			$\sin^2\theta^{l}_{23}$ 	& $0.3323$  		& $0.4207$\\
			\hline
			\hline
			$\chi^2_{\rm total}$	& $21.319\ $ & $15.222\ $\\ 
			\hline
			\hline
		\end{tabular}
		}
		}
	\caption{A sample pair of low-$\chi^2$ solutions in the $n=2$ case with 
	{$\eta\to 0$}
	(c.f. section \ref{sect:n2:54only}). 
		The four digit accuracy adopted in the physical parameters reflects the maximum quality of the input data these quantities are compared to, c.f. TABLE \ref{tab:inputs}. 
	Let us also remark that a full reconstruction of the displayed $\chi^{2}$ values an interested reader could attempt could be partly obscured by the limited precision of the displayed numbers.
	}
	\label{tab:BestSol_54only}
\end{table}

\subsubsection{Complete fits including $45_{H}$ \label{sect:54plus45}}
\begin{figure}[h]
\includegraphics[width=8.6cm]{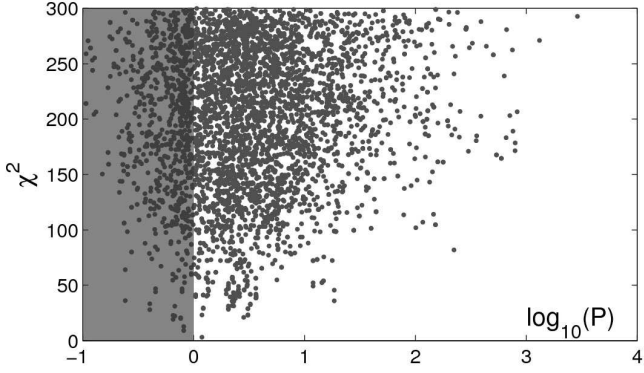}
\caption{{A sample of the $\chi^{2}$ values for the full $n=2$ fits as a function of the decoupling parameter $P$. One can see clearly that the extra constraints from the leptonic sector make the $n=2$ fits troublesome, see also section~\ref{sect:54plus45}.\label{fig:chisqxP45}}}
\end{figure}

Turning on the antisymmetric Yukawa coupling of the $45_{H}$ one could expect that the extra parameters associated to this sector would make the global fits of the measured quark and lepton masses and mixing parameters much simpler than in the 
{$\eta\to 0$
case discussed above. On the other hand, it is also clear that $45_{H}$ should have almost no impact in the quasi-decoupled regime when, effectively, only one of the two $10_{M}$'s contributes to the light states.}
Thus, the situation is more subtle and, as one can see in FIG.~\ref{fig:chisqxP45}, the extra $45_{H}$-term in the mass {formulas} actually leads to a significant improvement of the fits only in the very-low-$P$ region of the parametric space, c.f. FIG.~\ref{fig:chisqxP}. As before,  a detailed information about a pair of {low-$\chi^{2}$} solutions
can be found in TABLE~\ref{tab:BestSol_45on}. Indeed, since $P$ and $p$ are again strongly correlated, c.f. section~\ref{sect:54only}, in both cases also the $p$ parameter falls into the ${\cal O}(1)$ domain. 
\begin{table}[ht]
	\centering
		\begin{tabular}{|c|c|c|}
			\hline
			Parameter & Fit I & Fit II \\
			\hline
			\hline
			$p$	& $3.041675$ & $2.847552$ \\ 
			\hline\hline 
			$r$ & $0.01312141$ & $0.01499823$ \\ 
			$t$ & $87.744176$ & $162.609350$ \\ 
			\hline
			$\tilde{m}_u$ & $0.0003216498$ & $0.0003551049$ \\
			$\tilde{m}_c$ & $0.1461787$ & $0.1823098$ \\
			$\tilde{m}_t$ & $80.619489$ & $77.955444$ \\ 
			\hline
			$\tilde{F}_{11}$ & $ 0.006594 - 0.012611i$ & $-0.397570 - 0.498592i$ \\
			$\tilde{F}_{21}$ & $ 1.868096 + 0.406222i$ & $-0.320084 - 0.581020i$ \\
			$\tilde{F}_{31}$ & $-0.865185 + 12.270027i$ & $ 4.510621 - 1.124757i$ \\ 		
			$\tilde{F}_{12}$ & $ 0.192324 + 0.789105i$ & $ 0.976804 - 0.214744i$ \\ 
			$\tilde{F}_{22}$ & $ 2.488878 - 0.413259i$ & $ 3.817823 + 0.370953i$ \\ 
			$\tilde{F}_{32}$ & $ 0.321369 + 2.679685i$ & $ 7.491695 - 3.380059i$ \\ 			
			\hline
			$c_{11}$ & $-7.019216 + 4.756008i$ & $-0.766062 - 1.508434i$ \\
			$c_{22}$ & $-2.349768 - 0.927223i$ & $-0.032584 + 0.034197i$ \\
			$c_{12}$ & $0.2841363 - 0.567570i$ & $ 0.769877 - 1.705836i$ \\
			\hline
			$d_{\Delta}$ & $-0.988269 + 0.783068i$ & $ 0.121933 + 0.169316i$ \\
			$d_{\Lambda}$ & $ 0.005013 + 0.001892i$ & $ 0.015713 + 0.000770i$ \\
			\hline
			\hline
			$m_u$ [MeV]   						& $0.4503$ 			& $0.4971$\\
  		$m_c$ [MeV]   						& $204.7$ 			& $255.2$\\
  		$m_t$ [GeV]   						& $112.9$ 			& $109.1$\\
  		\hline
			$m_d$ [MeV]   						& $0.6364$ 			& $1.0176$\\
  		$m_s$ [MeV]   						& $26.97$ 			& $30.79$\\
  		$m_b$ [GeV]   						& $1.268$ 			& $1.651$\\
  		\hline
   		$\sin\theta^{q}_{12}$ 		& $0.2241$ 			& $0.2248$\\
   		$\sin\theta^{q}_{23}$ 		& $0.03465$ 		& $0.03289$\\
  		$\sin\theta^{q}_{13}$ 		& $0.003243$ 		& $0.003546$\\
  		$\delta_\mathrm{CP}^{q}$  & $47.77^\circ$ & $55.13^\circ$\\
			\hline
			\hline
			$m_{e}$ [MeV] 						& $0.3562$ 			& $0.3571$\\
  		$m_{\mu}$ [MeV] 					& $75.30$ 			& $75.30$\\
 			$m_{\tau}$ [GeV] 					& $1.619$ 			& $1.607$\\
 			\hline
  		$\frac{\Delta m^2_{21}}{|\Delta m^2_{31}|}$ & $0.03226$  & $0.03226$\\
  		\hline
 			$\sin^2\theta^{l}_{12}$ 	& $0.3039$ 			& $0.3048$\\
			$\sin^2\theta^{l}_{23}$ 	& $0.4769$  		& $0.4152$\\
			\hline
			\hline
			$\chi^2_{\rm total}$			& $3.203\ $ 		& $9.187\ $\\ 
			\hline
			\hline
			{$\sin\theta^l_{13}$} 		& $0.269$ 			& $0.255$ \\
			\hline
			{$\delta_{CP}^{l}$} & $-10.57^\circ$ & $11.99^\circ$\\
			\hline
		\end{tabular}
\caption{A sample pair of low-$\chi^2$ solutions obtained in section~\ref{sect:54plus45} for the $n=2$ case.  
	Here $d_{\Delta,\Lambda}=t^{-1}(v_d^{16}/m_b)(V^{45}_{\Delta,\Lambda}/V^{16})\eta_{12}$ and the last two rows represent the relevant predictions for $\theta_{13}^{l}$ and $\delta_{CP}^{l}$ obtained with the corresponding fits, c.f. also FIG.~\ref{fig:theta13lept} and FIG.~\ref{fig:DiracCPphase}.}   
	\label{tab:BestSol_45on}
\end{table}
%
\vskip 2mm\paragraph{{Sterility of $45_{H}$ for $p\gtrsim 10$:}}\mbox{}\\
Although the tight link between $p$ and $P$ emerging in the $|c_{kl}|\sim {\cal O}(1)$ regime justifies the high degree of sterility of the $45_{H}$ contribution for large $p\sim P\gtrsim {\cal O}(100)$ values corresponding to a quasi-decoupling of the second $10_{M}$, it could be rather surprising that very good fits can be obtained only for $p\sim{\cal O}(1)$. 
A thorough inspection of the role of $45_{H}$ in the relevant mass formulas given in Appendix \ref{appendix} reveals that this is namely due  to the antisymmetry of the corresponding Yukawa coupling $\eta$ which gives rise to, e.g., a further ${\cal O}(m_{s}/m_{b})$ suppression of the $45_{H}$ effects in some of the quark sector observables, in particular the first and second generation masses and the 13 and 23 CKM mixing angles. 
\vskip 2mm\paragraph{Genuine predictions for $\theta_{13}^{l}$ and $\delta_{CP}^{l}$:}\mbox{}\\
In the fits above, we let only the well measured quark and lepton masses and mixing parameters contribute to the global $\chi^{2}$ function. The other observables, in particular the reactor mixing angle and the {leptonic CP phases} were left apart as genuine predictions of the current scheme. Indeed, for any specific fit, these can be calculated in terms of the other parameters listed in TABLE~\ref{tab:BestSol_45on}.
  
\begin{figure}[h]
\includegraphics[width=8.6cm]{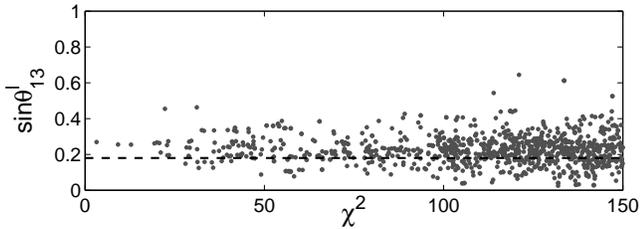}
\caption{{The predicted value of the leptonic $13$ mixing as a function of the $\chi^{2}$ corresponding to the fits of all the other measured parameters. The current
{90\% C.L.}
upper limit $\sin\theta_{13}^{l}\leq {0.18}$ is indicated by the dashed line. The distribution of the calculated $\theta_{13}^{l}$ values for the lowest-$\chi^{2}$ points clusters in the lower part of the available domain at around $\sin\theta_{13}^{l}\sim 0.2$. \label{fig:theta13lept}}}
\end{figure}

In FIG.~\ref{fig:theta13lept} we display the predicted values of the leptonic 13 mixing obtained for the {fits indicated in {FIG.~\ref{fig:chisqxP45}} with $\chi^{2}\lesssim 150$. Although it is impossible for the best-$\chi^{2}$ points to get within the current
{90\% C.L.}
experimental limit}, there is a {clear} preference of a small 1-3 mixing with $\sin\theta_{13}^{l}\sim 0.2$ at the low-$\chi^{2}$ tail of the distribution. 

\begin{figure}[h]
\includegraphics[width=8.5cm]{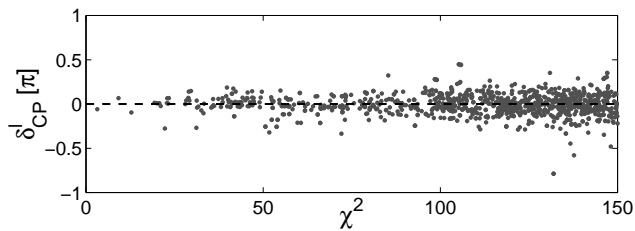}
\caption{{The predicted value of the leptonic Dirac CP phase $\delta^{l}_{CP}$ (in units of $\pi$) as a function of the $\chi^{2}$ corresponding to the fits of all the other measured parameters.  The distribution of the calculated $\delta^{l}_{CP}$ values for the lowest-$\chi^{2}$ points clusters at around zero, thus indicating a possible difficulty in revealing the leptonic CP violation in the next generation of neutrino oscillation experiments. \label{fig:DiracCPphase}}}
\end{figure}

{Similarly, as one can see in FIG.~\ref{fig:DiracCPphase}, a small leptonic Dirac CP phase {$\delta_{CP}^{l}$} is strongly preferred for the {lowest-$\chi^{2}$} solutions.}
{As far as the Majorana phase is concerned (recall that one of the light neutrinos is exactly massless in the current setting) we do not observe any specific feature in its distribution and the predictions are essentially uniformly covering the whole available domain.} 
\vskip 2mm
\subsection{More than two vector matter multiplets?\label{sec:threeandmore}}
As we have seen, even with 2 copies of extra matter multiplets at play the fits of the system (\ref{calMd2})-(\ref{calMnu2}) are {non-trivial, although a na\"ive parameter counting (see section~\ref{sect:parametercounting}) would clearly suggests the opposite. As a matter of fact, this is very welcome because such a setting admits to draw genuine prediction that can be tested at near-future experimental facilities.} 

{From an underlying $E_{6}$ perspective one could ask whether a third $10_{M}$ would cause a qualitative change of the picture}. Given the number of extra parameters popping up in the $n=3$ setting the general answer is very likely to be positive. For the same reason, this is not the strategy we would like to pursue as it would most probably lack any predictive power. Moreover, reiterating the hierarchy arguments given in section~\ref{sect:introduction}, the overall scale of the corresponding $M_{10}$ would be unnaturally low if one brought all three of its eigenvalues below the Planck scale. Apart from these rather technical issues, a third $10_{M}$ would not shed any new light onto the neutrino sector which, in spite of its appeal, doesn't need to be dominated by the triplet contribution at all. 

Hence, without a handle on the non-renormalizable terms governing the type-I sector, we consider further ($n\geq3$) extensions of the current analysis to be rather academic. Nevertheless, there are various concepts that can provide an extra information making such studies nontrivial and potentially interesting, be it family symmetries, extra dimensions, finite unifications or anything else.
This, however, is beyond the scope of this study.
 
\section{Conclusions and outlook}
In this work we have studied in detail the flavour structure of the simple SUSY $SO(10)$ GUT models with extra 10-dimensional vector multiplets admixing with the ``standard'' 16-dimensional matter spinors which provide an interesting link between the relative magnitude of the $SU(2)_{R}\otimes U(1)_{B-L}$ and $SU(5)$ breaking observed in the SM matter spectra and the hierarchy of the SUSY GUT-scale thresholds. We argued that this setting is very well motivated if, for instance, the flavour structure of the gauge-singlet mass term of the matter 10's exhibits a few-orders-of-magnitude hierarchy. Moreover, this class of models received a further credit in the recent works \cite{gaugemediation} where it was shown to be capable of accommodating a simple tree-level realisation of the gauge-mediated supersymmetry breaking mechanism in a phenomenologically viable manner.  

Focusing on the next-to-minimal case with more than a single such matter 10 playing an active role in the effective SM matter spectrum, the well-known no-go emerging in the minimal model already at the charged sector level is alleviated. Subsequently, we have pursued an extensive $\chi^{2}$ analysis including the neutrino sector data which is tractable only if the calculable $SU(2)_{L}$-triplet contribution dominates the neutrino mass matrix. 
{We have obtained very good fits of all quark and lepton masses and mixing parameters measured so far, providing a pair of genuine predictions for those to be, presumably, within the reach of the near future facilities: the reactor mixing angle is predicted to be relatively large, close to the current {$90\%$ C.L. limit quoted, e.g., in \cite{theta13}}, while the leptonic Dirac CP phase tends to be very small and, hence, more difficult to access. {No preference has been observed for the relevant Majorana CP phase.}}  

{
Unfortunately, a full account of the neutrino sector including also the type-I-like contribution to the seesaw formula is intractable without an additional information on how the full rank of the heavy part of the neutrino mass matrix is restored in a specific setting. For instance, one can think about extra flavour symmetries \cite{Froggatt:1978nt} that may, at least to some extent, keep the number of free parameters under control, and at the same time constrain the Yukawa couplings of the model. 
This, however, typically requires non-renormalizable operators to be invoked at some level, thus challenging the original motivation for the extra vector-like matter as a renormalizable key to the observed quark and lepton masses and mixing. Nevertheless, at closer look such models are likely to exhibit rather different correlation patterns due to the generic dominance of better-controlled renormalizable contributions. In this respect, an extra flavour symmetry could be at least partially unloaded from the usual burden of having to address many aspects of the SM flavour problem at once, as it is often required when matter is spanned over the $SO(10)$ spinors only.
}

\section*{Acknowledgments}
The work of M. M. was supported by the Royal Institute of Technology (KTH), Contract No. SII-56510, by a Marie Curie Intra European Fellowship within the 7th European Community Framework Programme FP7-PEOPLE-2009-IEF, contract number PIEF-GA-2009-253119,  by the EU Network grant UNILHC PITN-GA-2009-237920, by the Spanish MICINN grants FPA2008-00319/FPA and MULTIDARK CAD2009-00064 (Con-solider-Ingenio 2010 Programme) and by the Generalitat Valenciana grant Prometeo/2009/091.  The work of M.H. has been supported by the Studienstiftung des deutschen Volkes (SdV) and by the Deutscher Akademischer Austausch Dienst (DAAD). He warmly acknowledges the hospitality of the Particle theory group of the KTH Theoretical physics department.

\vskip 10mm
\appendix
\section{Sterility of $45_{H}$ for $p\gtrsim 10$\label{appendix}}
As seen in section~\ref{sect:n2:54only}, in the $\eta\to 0$ limit the good fits of all observables including the neutrino ones required a very mild hierarchy between the two eigenvalues of $M_{10}$, i.e., $p \lesssim 10$. Remarkably, such a strong preference of very low $p$ values appears also in the general fits with non-zero $\eta$ where one would expect it to be much weaker due to the extra freedom associated to the active role of $\vev{45_{H}}$ in the relevant formulas. 

Although it is quite difficult to provide a general understanding of this behaviour one can address at least some of its aspects. In particular, one can decipher why for a given moderate-$\chi^{2}$ point obtained in the $54_{H}$-only fits of section~\ref{sect:54only} the extra freedom associated to the subsequent inclusion of $45_{H}$ in section~\ref{sect:54plus45} does not improve the $\chi^{2}$ of the corresponding complete fits.

First, it is easy to show that for $2\times 2$ matrices the antisymmetric\hyphenation{anti-sym-met-ric} part of the inverse of an arbitrary matrix $M$ is a function of the antisymmetric part of $M$ only, apart from an overall normalization. Note that this specific to $2\times 2$ matrices and does not hold for larger dimensionality. 
Thus, the antisymmetric $\tilde\eta$ enters the formulas (\ref{calMd2})-(\ref{calMe2}) in a very specific manner: it only generates an extra antisymmetric contribution to the ubiquitous symmetric part of the $\tilde{F}\tilde{M}^{-1}\tilde{F}^{T}$ bilinears generated by the $\tilde{M}_{10}$- and $\tilde\lambda$- pieces in (properly normalized) eqs.~(\ref{MDeltaLambda}). As such, barring the sub-leading effect it has in $H_{d}$ and $H_{e}$, it can affect the relevant mixing angles at the ${\cal O}(\varepsilon)$ level while the spectrum of the matrices (\ref{calMd2})-(\ref{calMe2}) remains intact up to ${\cal O}(\varepsilon^{2})$ corrections~\cite{Bertolini:2004eq} where $\varepsilon$ is parametrizing the ``smallness'' of the antisymmetric correction as compared to the symmetric one given by the remaining terms in (\ref{calMd2})-(\ref{calMe2}). Since the $\tilde{F}\tilde{M}_{\Delta,\Lambda}^{-1}\tilde{F}^{T}$ bilinears are tailored to give rise to the second generation masses, the typical size of their leading order entries in the down-quark sector is $m_{s}/m_{b}$ while it is $m_{\mu}/m_{\tau}$ for the charged leptons, both in the few percent range. However, this is all namely due to its symmetric part dominated by $\tilde{M}_{10}$ and for $p>1$ there is an extra overall suppression associated to the antisymmetric piece. This can be seen, e.g.,  from 
\be
(\tilde M_{\Delta})^{-1}=\frac{1}{{\rm det} \tilde M_{\Delta}}\left[\begin{pmatrix}s_{22} & -s_{12} \\ -s_{12} & s_{11}\end{pmatrix}+\begin{pmatrix}0 & a \\ -a & 0\end{pmatrix}\right],
\ee 
where $s_{kl}= (\tilde{M}_{10}-\tilde \lambda)_{kl}$, $a= \tilde{\eta}_{12}$
 and ${\rm det} \tilde M_{\Delta}=s_{11}s_{22}-s_{12}^{2}+a^{2}$. It is clear that for moderate $\tilde{\lambda}$ and $\tilde{\eta}$ of the order of ${\cal O}(t)$ the leading contribution to the symmetric part of $\tilde{F}\tilde{M}^{-1}\tilde{F}^{T}$ scales as $|\tilde F|^{2}s_{kk}/s_{11}s_{22}\approx |\tilde F|^{2}/t$ while the antisymmetric piece $|\tilde F|^{2}a/s_{11}s_{22}\approx |\tilde F|^{2}/pt$ is suppressed by an extra factor of $p^{-1}$. Thus, the relevant $\varepsilon$ parameter behaves like $m_{s}/m_{b}\, p \sim 0.02/p$.
From this, it is already clear that {for the fits with $p\gtrsim 10$} such an antisymmetric correction can not help lowering the $\chi^{2}$ value of a specific fit if it comes predominantly from the second generation of down quark or charged lepton masses.

Concerning the impact of a non-negligible $\eta$-contribution to the mixing parameters the situation is somewhat more subtle.   
At the leading order, one can quantify the shift in the CKM mixing angles as:
\be
V_{CKM}' \approx V_{CKM} (1-Z) \,,
\ee
where $Z$ is an anti-hermitean matrix obtained from
$$
Z_{ij,i<j}\approx{A_{ij}}/{(S_{d})_{jj}} 
$$
and $A$ stands for the antisymmetric part of $\tilde{F}(\tilde{M}_{\Delta})^{-1}\tilde{F}^{T}$ in the basis in which the symmetric part of ${\cal M}_{d}$, $S_{d} \approx \left( r \tilde{\cal M}_u - \tilde{F} (\tilde{M}_\Delta)^{-1} \tilde{F}^T \right)$, is diagonal and real. Since, as we have seen,  $A$ is $1/p$-suppressed with respect to the symmetric part of $\tilde{F}(\tilde{M}_{\Delta})^{-1}\tilde{F}^{T}$, one has
\be
Z \approx \frac{1}{p}\begin{pmatrix}
0 & {\cal O}(1) &{\cal O}(\delta) \\
. & 0 &{\cal O}(\delta)  \\
. & . & 0
 \end{pmatrix} \quad {\rm with} \quad \delta \equiv m_{s}/m_{b}\,,
\ee
where we have approximated the eigenvalues of $S_{d}$ by $\{m_{d}/m_{b},m_{s}/m_{b},1\}$.
{Thus, for the ``transition region'' values of the $p$-parameter, i.e., {$p\gtrsim 10$} under consideration, only $\theta_{12}^{q}$ can be slightly} affected by the Yukawa of $45_{H}$ while the other CKM mixings remain essentially intact. This, however, does not improve the fits with {``moderate'' $\chi^{2}$ value which is spanning over several different observables other than just $\theta_{12}^{q}$.}

{Let us also remark that the situation in the leptonic sector is very similar to quarks, in particular for the charged lepton contribution to the leptonic mixing. Moreover, the current precision of the leptonic mixing parameters determination is much worse than the same in the quark sector so the net effect of the antisymmetric Yukawa of $45_{H}$ in the leptonic mixing $\chi^{2}$ contribution is essentially negligible for $p\gtrsim 10$.}

\end{document}